\documentclass[12pt]{article}
\usepackage[dvips]{color}
\usepackage{epsfig}
\usepackage{amsmath}
\usepackage{graphicx}

\begin{document}
\title{\bf Quantum loop corrections of charged dS black hole}
\author{{J. Naji\thanks{Email: naji.jalil2020@gmail.com}}\\
{\small {\em Physics Department, Ilam University, Ilam, Iran}}\\
{\small {\em P.O. Box 69315-516, Ilam, Iran}}} \maketitle
\begin{abstract}
\noindent In this paper, a charged black hole in de Sitter space considered and logarithmic corrected entropy used to study thermodynamics. Logarithmic corrections of entropy comes from thermal fluctuations which play role of quantum loop corrections. In that case we are able to study the effect of quantum loop on the black hole thermodynamics and statistics. As black hole is a gravitational object, so it helps to obtain some information about the quantum gravity.  The first law of thermodynamics investigated for logarithmic corrected case and find that is only valid for the charged dS black hole. We show that the black hole phase transition disappear in presence of logarithmic correction.

\end{abstract}
\section{Introduction}
As we know, important black hole quantities like horizon radius, mass, charge, temperature and so on are related by an equation of the form the first law of thermodynamics \cite{1}. Therefore, the black hole thermodynamics is important and interesting field of research. Among several kind of black hole, study of asymptotically anti-de Sitter (AdS) black holes are interesting from aspects of AdS/CFT \cite{2} or AdS/QCD correspondence \cite{3,4}. In another hand, there is a correspondence between a gravitational theory in de Sitter space and conformal field theory \cite{5}. In that case thermodynamics study of asymptotically de Sitter (dS) space is interesting and important for example see Ref. \cite{dS}. dS black holes are also interesting from cosmological point of view. Because in addition at the early universe, during inflationary epoch universe was a de Sitter space, and in far future it will turn into a de Sitter space \cite{ds2,ds3}. Therefore, we need to formulate  the dS/CFT correspondence to apply gauge/gravity duality to the universe. For example, Ref. \cite{6} similar to the AdS case, proposed a new method to study thermodynamics of dS black holes. They found information about possible phase transitions in various dimensions. The main goal of our paper is to consider a charged dS black hole and study thermodynamics properties under thermal fluctuation effects. Thermal fluctuations are small perturbation about equilibrium so such analysis is valid at the first order approximation. Such thermal fluctuations, which are important when size of black hole is small, play roll of quantum loop correction and show itself by logarithmic correction in the entropy of the black hole. Such logarithmic corrected entropy proposed by the Refs. \cite{7, 8}. In the Ref. \cite{7} leading-order corrections to the entropy of any thermodynamic system due to small statistical fluctuations around equilibrium computed and show that is logarithmic. In the Ref. \cite{8} a charged black hole with a scalar field which is coupled to gravity in (2 + 1)-dimensions \cite{9, 10, 11, 12, 13} considered and logarithmic corrections to the corresponding system using two approaches computed. In that case the effects of thermal fluctuations on a charged AdS black hole has been investigated by \cite{14}. Interestingly, thermodynamics and statistics of G\"{o}del black hole with logarithmic correction has been studied by the Ref. \cite{15}.  This method is applicable for various kinds of black objects such as black Saturn \cite{16} or charged dilatonic black Saturn \cite{17}. Also the effects of thermal fluctuations on the thermodynamics of a modified Hayward black hole has been considered by the Ref. \cite{18}. In the other works P-V criticality of charged AdS space-time \cite{momen} using AdS/CFT  and dyonic charged AdS black hole \cite{19} which is holographic dual of a Van der Waals fluid \cite{20} considered and logarithmic correction analyzed. In all above works quantum gravitational effects has been studied using black objects.
In this work we would like to consider a charged dS black hole with cosmological constant and study thermodynamics under effect of thermal fluctuations which is indeed quantum loop effect. We study this effect in the black hole phase transition.
This paper is organized as follows. In the next section we review some important properties of charged dS black hole. In the section 3 we study thermodynamics with logarithmic corrected entropy. In the section 4 we study phase transition and finally in section 5 we give conclusion.

\section{Charged dS black hole}
\begin{figure}[h!]
 \begin{center}$
 \begin{array}{cccc}
\includegraphics[width=50 mm]{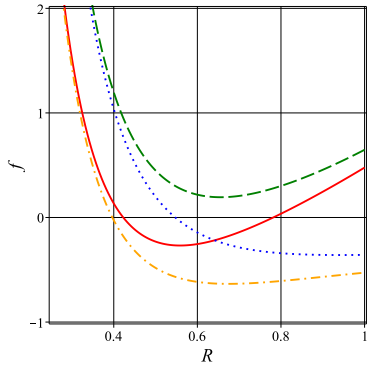}
 \end{array}$
 \end{center}
\caption{$f$ in terms of $R$ for $b=0.1$ and $a=0.8$ (solid red),  $b=0.1$ and $a=0.9$ (dash green),  $b=-0.02$ and $a=0.9$ (dot blue), $b=-0.02$ and $a=0.8$ (dash dot orange).}
 \label{fig1}
\end{figure}

In this section we review some important properties of Charged dS black hole which will be useful for our future aims.
The charged dS black hole in four dimensions given by the following line element \cite{6},
\begin{equation}\label{1}
ds^{2}=-fdt^{2}+\frac{dr^{2}}{f}+r^{2}d\Omega_{2}^{2},
\end{equation}
with
\begin{equation}\label{2}
f=1-\frac{2M}{r}+\frac{Q^{2}}{r^{2}}+\frac{8\pi P r^{2}}{3},
\end{equation}
where $M$ and $Q$ denote black hole mass and charge respectively and pressure $P$ related to the cosmological constant which has negative value for dS and positive value for AdS black hole. $d\Omega_{2}^{2}$ shows the line element of the unit 2-sphere. The metric (\ref{1}) reduced to Schwarzschild-dS black hole for $Q=0$. In order to obtain thermodynamics relations, it is important to have black hole horizon radius which is obtained from $f=0$. Using appropriate choice of black hole parameters $M$, $P$ and $Q$ we can find two real positive roots of $f=0$. In the Fig. \ref{fig1} we draw $f$ in terms of dimensionless quantities,
\begin{equation}\label{2}
R=\frac{r}{M},\hspace{1cm} a=\frac{Q}{M},\hspace{1cm} b=PM^{2},
\end{equation}
and we can see that there is a minimum which means $f^{\prime}=0$ and $f^{\prime\prime}>0$, where prime is derivative with respect ro $r$. Therefore, there are two real and positive roots $r_{\pm}$ corresponding to inner and outer horizons.

One can write the black hole mass in terms of outer horizon radius as,

\begin{equation}\label{3}
M=\frac{r_{+}}{2}+\frac{Q^{2}}{2r_{+}}+\frac{4\pi P r_{+}^{3}}{3}.
\end{equation}

Minimum in the Fig. \ref{fig1} obtained by $f^{\prime}(r_{m})=0$ where $r_{m}$ is location of minimum with the following condition $r_{-}\leq r_{m}\leq r_{+}$. Special case of $r_{-}= r_{m}= r_{+}\equiv r_{e}$ is corresponding to the extremal black hole with zero temperature. In that case one can obtain the following relation,

\begin{equation}\label{4}
Q^{2}=r_{e}^{2}(8\pi P r_{e}^{2}+1).
\end{equation}
It means that,
\begin{equation}\label{5}
r_{e}=-\frac{\sqrt{-\pi P(1+\sqrt{32\pi P Q^{2}})}}{4\pi P}.
\end{equation}
One can use above relations to study extremal limits.\\
For the ordinary charged dS black hole, corresponding thermodynamics quantities given by the following equations. Hawking temperature given by,
\begin{equation}\label{6}
T=\frac{1}{4\pi r_{+}}(1+8\pi P r_{+}^{2}-\frac{Q^{2}}{r_{+}^{2}}),
\end{equation}
while black hole volume reads,
\begin{equation}\label{7}
V=\frac{4}{3}\pi r_{+}^{3},
\end{equation}
and chemical potential corresponding to the black hole charge given by,
\begin{equation}\label{8}
\Phi=\frac{Q}{r_{+}}.
\end{equation}
Then, one can obtain specific heat and see that there are some instabilities which discussed in details by the Ref. \cite{6}. We also discuss about that and show that such phase transition disappear in presence of logarithmic corrections.

\section{Logarithmic corrected thermodynamics}
\begin{figure}[h!]
 \begin{center}$
 \begin{array}{cccc}
\includegraphics[width=50 mm]{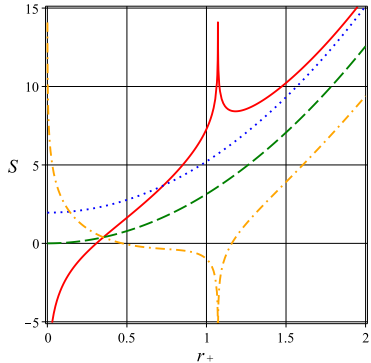}
 \end{array}$
 \end{center}
\caption{$S$ in terms of $r_{+}$ for $P=-0.0045$. $Q=1$ and $\alpha=0$ (dash green), $Q=1$ and $\alpha=0.5$ (solid red), $Q=0$ and $\alpha=0.5$ (dotted blue), $Q=1$ and $\alpha=-0.5$ (dash dot orange).}
 \label{fig2}
\end{figure}

The entropy of a black hole can correct by a logarithmic term because of the thermal fluctuations. These are important as the size of black hole reduces due to the Hawking radiation. It is possible to relate the microscopic degrees of freedom of a black hole with a conformal field theory.
In that case the modular invariance of the partition function can constrain the entropy of the charged dS black hole.\\
One can calculate a field configuration amplitude by using the path integral formalism  to propagate to another field configuration. It can performed by using the Euclidean quantum gravity method. Thus, the partition function for the charged dS black hole can be written as \cite{02ab},
\begin{equation}
Z = \int D g  D A e^{- I},
\end{equation}
where $I \to -i I$ is the Euclidean action of the system.
The above partition function may related to the statistical mechanics partition function given by \cite{hawk, hawk1},
\begin{equation}
Z = \int_0^\infty  dE \, \,  \rho (E) e^{-\beta E},
\end{equation}
where $\beta$ denotes the inverse of the temperature in the units of Boltzmann constant.
Hence, we can write the density of states as follow,
\begin{eqnarray}
\rho (E) = \frac{1}{2 \pi i} \int^{\beta_0+ i\infty}_{\beta_0 - i\infty}  d \beta \, \, e ^{S(\beta)} ,
\end{eqnarray}
where
\begin{equation}
S = \beta  E   + \ln Z.
\end{equation}
It is usual that the entropy measured around the equilibrium $\beta_0$, and all thermal fluctuations are
neglected. However, it is possible to consider mentioned thermal fluctuations and expand $S(\beta)$ around the equilibrium temperature $\beta_0$ \cite{7},
\begin{equation}\label{fluc}
S = S_0 + \frac{1}{2}(\beta - \beta_0)^2 \left(\frac{\partial^2 S(\beta)}{\partial \beta^2 }\right)_{\beta = \beta_0},
\end{equation}
where $S_0$ is uncorrected entropy and the higher order corrections of the entropy neglected. Therefore, one can write the density of states as follow,
\begin{eqnarray}
\rho (E) = \frac{e^{S_0}}{ 2 \pi i}  \int^{\beta_0+ i\infty}_{\beta_0 - i\infty}  d \beta \, \,  \exp \left( \frac{1}{2}
 (\beta- \beta_0)^2 \left(\frac{\partial^2 S(\beta)}{\partial \beta^2 }\right)_{\beta = \beta_0}   \right).
\end{eqnarray}
By using a change of variable one can obtain,
\begin{equation}
\rho(E) = \frac{e ^{S_{0}}}{\sqrt{2 \pi }} \left[\left(\frac{\partial^2 S(\beta)}{\partial \beta^2 }\right)_{\beta = \beta_0}\right]^{1/2}.
\end{equation}
Hence one can write,
\begin{equation}
S = S_0 -\frac{1}{2} \ln \left[\left(\frac{\partial^2 S(\beta)}{\partial \beta^2 }\right)_{\beta = \beta_0}\right]^{1/2}.
\end{equation}
We can say that the second derivative of entropy is indeed the squared of the energy fluctuation.
One can rewrite this expression by using the fact that the microscopic degrees of freedom of a charged dS black hole may be calculated
using a conformal field theory \cite{card}. Therefore, the entropy can be written as,
\begin{equation}
S = a \beta^m  + b \beta^{-n },
\end{equation}
where $a$, $b$, $m$, and $n$ are positive constants \cite{ca}. The extremum value obtained by $\beta_0 = (n b/m a)^{1/(m+n)}$, and so expanding $S$ around the extremum yields to the following expression,
\begin{equation}
 \left(\frac{\partial^2 S(\beta)}{\partial \beta^2 }\right)_{\beta = \beta_0}  = S_0 \beta_0^{-2}.
\end{equation}
Finally, we use the corrected form of the entropy as
\begin{equation}\label{9}
S = \pi r_{+}^{2} -\alpha \ln (\pi r_{+}^{2} T^{2}),
\end{equation}
where $S_{0}=\pi r_{+}^{2}$ is used. In the above relation, $\alpha=\frac{1}{2}$ usually used for the corrected case while $\alpha=0$ corresponding to the usual thermodynamics with the entropy $S_{0}=\pi r_{+}^{2}$. However, other values of $\alpha$ is possible.\\
In the Fig. \ref{fig2} we can see the effect of logarithmic correction on the black hole entropy. We draw typical behavior for the specific choice of $Q$ and $P$, however general behavior is the same for another choices instead of $Q=0$. In the case of $\alpha=0$ we can see that the entropy is completely increasing function. In the case of $\alpha\neq0$ we find that there is two critical horizon radius. At the first critical point $r_{+c1}$ we can see asymptotic like behavior.  Before that point the entropy is totally increasing function. After that it decreases to the second critical point $r_{+c2}$ which is an extremum where $\frac{dS}{dr_{+}}=0$. Then, again the entropy is increasing function and the second law of thermodynamics is satisfied. These tow critical points are local maximum and minimum of the entropy. We can interpret the second extremum as the possible smallest radius of black hole. It means that the black hole with $r_{+}\leq r_{+c2}$ can not exist, however this smallest value of $r_{+}$ may be satisfy $S(r_{+0})=0$. It is clear that $r_{+c2}>r_{+c1}>r_{+0}$. The second critical point is real positive root of the following equation,
\begin{equation}\label{10}
16\pi^{2}Pr^{6}+2\pi(1-16\alpha P)r^{4}-2\pi Q^{2}r^{2}-4\alpha Q^{2}=0.
\end{equation}
For the large $r_{+}$, as expected, the effect of logarithmic correction is negligible and importance of thermal fluctuations is for small $r_{+}$.\\

Assuming constant $Q$ and $P$, we can investigate the first law of thermodynamics,
\begin{equation}\label{11}
dM=TdS+\Phi dQ+VdP.
\end{equation}
Clearly, above equation satisfied for $\alpha=0$ but in order to satisfy for the logarithmic correction we should have,
\begin{equation}\label{12}
r_{+}=\left(-\frac{Q^{2}}{8\pi P}\right)^{\frac{1}{4}}.
\end{equation}
Validity of the first law of thermodynamics is only for the negative pressure corresponding to the charged dS black hole. For the selected value of $P=-0.0045$ and $Q=1$ corresponding to the Fig. \ref{fig2} we have $r_{+}=1.7244$. Again we mention that these are typical values, and we can obtain similar results with other values.\\
Black hole mass treated as the gravitational enthalpy, so,

\begin{equation}\label{13}
H=\frac{r_{+}}{2}+\frac{Q^{2}}{2r_{+}}+\frac{4\pi P r_{+}^{3}}{3}.
\end{equation}

Then, Gibbs free energy is given by the Legendre transformation of the enthalpy,

\begin{eqnarray}\label{14}
G&=&H-TS\\
&=&\frac{2}{\pi {r_{+}}^{3}}\left( P\pi {r_{+}}^{4}-\frac{{Q}^{2}}{8}+\frac{{r_{+}}^{2}}{8} \right)
\alpha\,\ln  \left( {\frac { \left(
8\,P\pi \,{r_{+}}^{4}-{Q}^{2}+{r_{+}}^{2} \right)^{2}}{16\pi \,{r_{+}}^{4}}}\right)\nonumber\\
&+&\frac{1}{12}\left( -8\,P{\pi }{r_{+}}^{3}+9\, \,{Q}^{2}{r_{+}}^{-1}+3\, \,{r_{+}}\right).\nonumber
\end{eqnarray}

\begin{figure}[h!]
 \begin{center}$
 \begin{array}{cccc}
\includegraphics[width=50 mm]{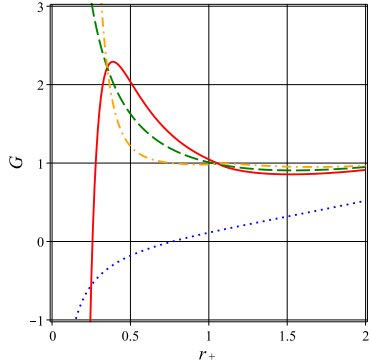}
 \end{array}$
 \end{center}
\caption{$G$ in terms of $r_{+}$ for $P=-0.0045$. $Q=1$ and $\alpha=0$ (dash green), $Q=1$ and $\alpha=0.5$ (solid red), $Q=0$ and $\alpha=0.5$ (dotted blue), $Q=1$ and $\alpha=-0.5$ (dash dot orange).}
 \label{fig3}
\end{figure}

In the Fig. \ref{fig3} we can see typical behavior of Gibbs free energy in terms of $r_{+}$ which show that at low temperature or large size of black hole, logarithmic corrections are negligible. As discussed by the Ref. \cite{6}, in the case of charged dS black hole with $\alpha=0$, there is no small/large black hole phase transition, and behavior of the Gibbs free energy is usual, while in presence of the logarithmic correction we can see such phase transition. Gibbs Free energy have negative infinite value for small black hole and yields to a positive constant for the large black hole. There is a maximum for critical value of the black hole horizon.

\begin{figure}[h!]
 \begin{center}$
 \begin{array}{cccc}
\includegraphics[width=50 mm]{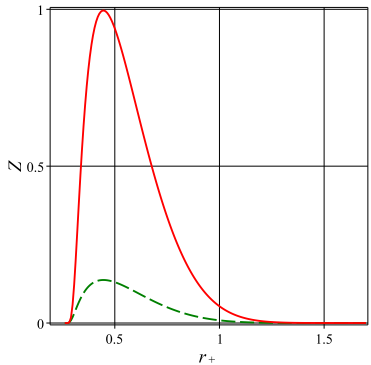}
 \end{array}$
 \end{center}
\caption{$Z$ in terms of $r_{+}$ for $P=-0.0045$ and $Q=0.26$.  $\alpha=0$ (dash green), $\alpha=0.5$ (solid red).}
 \label{fig4}
\end{figure}

We can also study behavior of Helmholtz free energy via,
\begin{equation}\label{15}
F=E-TS,
\end{equation}
where
\begin{equation}\label{15-1}
E=\int{TdS}=\frac{4}{3}\pi Pr_{+}^{3}-8\alpha Pr_{+}+\frac{r_{+}}{2}+\frac{Q^{2}}{2r_{+}}+\frac{\alpha Q^{2}}{3\pi r_{+}^{3}},
\end{equation}
is internal energy. Behavior of Helmholtz free energy under effect of thermal fluctuations is similar to the Gibbs free energy. Helmholtz free energy is important thermodynamics quantity to study stability of the black hole. We return to this point in the next section. Now we can use Helmholtz free energy to extract partition function of the canonical ensemble,
\begin{equation}\label{15-2}
Z=e^{-\frac{F}{T}}.
\end{equation}
In the Fig. \ref{fig4} we can see typical behavior of the partition function. It is clear that effect of the logarithmic correction is increasing of probability.

\section{Stability}
We will analyze the stability of the model using both specific heat and the entire Hessian matrix of the free energy. The last is important in any systems with chemical potential \cite{21}. First of all we should calculate specific heat,

\begin{equation}\label{16}
C=T(\frac{dS}{dT})=\frac{16\pi^{2}Pr_{+}^{6}+2\pi (1-16\alpha P) r_{+}^{4}-2\pi Q^{2}r_{+}^{2}-4\alpha Q^{2}}{8\pi P r_{+}^{4}+3Q^{2}-r_{+}^{2}}.
\end{equation}

\begin{figure}[h!]
 \begin{center}$
 \begin{array}{cccc}
\includegraphics[width=50 mm]{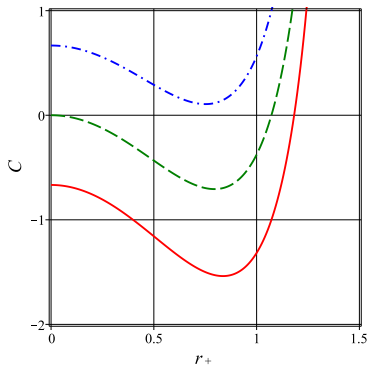}
 \end{array}$
 \end{center}
\caption{$C$ in terms of $r_{+}$ for $P=-0.0045$ and $Q=1$.  $\alpha=0$ (dash green), $\alpha=0.5$ (solid red), $\alpha=-0.5$ (dash dot blue).}
 \label{fig5}
\end{figure}

In the Fig. \ref{fig5} we can see behavior of the specific heat with variation of correction parameter. Surprisingly, we can see that negative value of $\alpha$ can remove instability of the black hole and hence there is no phase transition. Otherwise, usual value of $\alpha$ change smallest value of the black hole horizon. However we need more study to decide about black hole stability.

Because of presence of the black hole charge we can check above statements using the
matrix of second derivatives of free energy with respect to temperature $T$ and chemical
potential $\Phi$ given by the following components,
\begin{eqnarray}\label{17}
H_{11}&=& \frac{\partial^{2}F}{\partial T^{2}},\nonumber\\
H_{12}&=&\frac{\partial^{2}F}{\partial T\partial\Phi},\nonumber\\
H_{21}&=& \frac{\partial^{2}F}{\partial\Phi\partial T},\nonumber\\
H_{22}&=& \frac{\partial^{2}F}{\partial \Phi^{2}}.
\end{eqnarray}
It is easy to check that $H_{11}H_{22}-H_{12}H_{21}=0$, which means that one of the
eigenvalues is zero and we should seek the other which is the matrix trace given by,

\begin{equation}\label{18}
Tr(H)=H_{11} + H_{22}
\end{equation}

\begin{figure}[h!]
 \begin{center}$
 \begin{array}{cccc}
\includegraphics[width=50 mm]{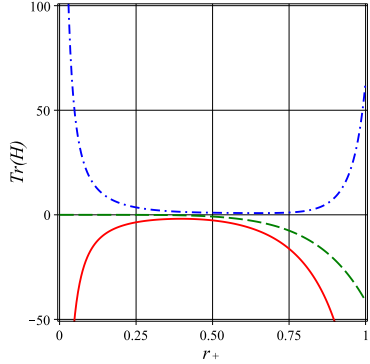}
 \end{array}$
 \end{center}
\caption{Hessian trace in terms of $r_{+}$ for $P=-0.0045$ and $Q=0.1$. $\alpha=0$ (dash green), $\alpha=0.5$ (solid red), $\alpha=-0.5$ (dash dot blue).}
 \label{fig6}
\end{figure}

In the Fig. \ref{fig6} we draw variation of Hessian trace with respect to the black hole horizon by variation of the correction parameters. In agreement with specific heat analysis we find that complete stability of the charged dS black hole obtained by negative $\alpha$.

\begin{figure}[h!]
 \begin{center}$
 \begin{array}{cccc}
\includegraphics[width=50 mm]{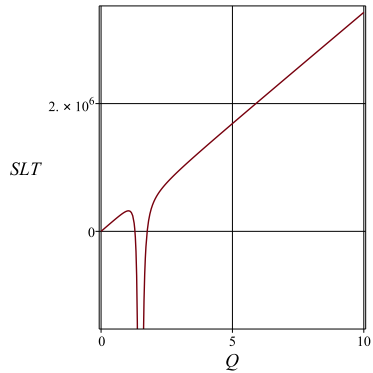}
 \end{array}$
 \end{center}
\caption{The second law of thermodynamics in terms of $Q$ for $P=-0.0045$.}
 \label{fig7}
\end{figure}

Therefore, we can use the following entropy to avoid phase transition of the charged dS black hole,
\begin{equation}\label{end}
S = \pi r_{+}^{2} +\frac{1}{2} \ln (\pi (\frac{1}{4\pi}(1+8\pi P r_{+}^{2}-\frac{Q^{2}}{r_{+}^{2}}))^{2}).
\end{equation}
Still, black hole charge is important parameter in study of phase transition. Therefore, we use relation (\ref{12}) in the equation (\ref{end}) and consider  geometro-thermodynamic formalism \cite{Geo2, Geo3} to investigate the first and the second law of thermodynamics simultaneously. In that case the thermodynamics metric given by,
\begin{equation}\label{Geo1}
g=\left(E^{a}\frac{\partial S}{\partial E^{a}}\right)\left(\eta_{b}^{c}\frac{\partial^{2}S}{\partial E^{c}\partial E^{d}}dE^{b}dE^{d}\right),
\end{equation}
where $\eta_{b}^{c}=(-1, 1, 1,\cdots, 1)$. We should note that $E^{a}$ are the relevant extensive parameters of the charged dS black hole which are $Q$ and $P$. Hence, the thermodynamic metric reduced to,
\begin{equation}\label{Geo2}
g=g_{11}dQ^{2}+g_{22}dP^{2}=\left(Q\frac{\partial S}{\partial Q}+P\frac{\partial S}{\partial P}\right)
\left(-\frac{\partial^{2}S}{\partial Q^{2}}dQ^{2}+\frac{\partial^{2}S}{\partial P^{2}}dP^{2}\right).
\end{equation}
Therefore, the second law of thermodynamics given by \cite{Geo4},
\begin{equation}\label{Geo4}
SLT\equiv\frac{\partial^{2}S}{\partial Q^{2}}+\frac{\partial^{2}S}{\partial P^{2}}\geq0.
\end{equation}
In the Fig. \ref{fig7} we study relation (\ref{Geo4}) for several values of $Q$ and find that equation (\ref{Geo4}) satisfied except for special values of $Q$. Situation is similar for other negative values of $P$. Hence it is find that the corrected entropy (\ref{end}) with small charge and negative pressure yields to stable black hole with valid first and second laws of thermodynamics.

\section{Conclusion}
In this work a charged black hole in de Sitter space with negative pressure considered. Thermodynamics of the model has been investigated under effect of logarithmic corrected entropy. Some important thermodynamics quantities calculated and effects of thermal fluctuations using logarithmic term in the entropy analyzed. The motivation to consideration of this system is the dS/CFT Correspondence \cite{22}. For example one can use results of this paper to study dark energy thermodynamics. Also, the production of charged pairs by a uniform electric field in a dS space may affected by logarithmic term \cite{Sh}.
First of all we reviewed some black hole properties like horizon structure and extremal case. Then, we considered logarithmic corrected entropy and found appropriate condition to satisfy the first law of thermodynamics. We studied thermodynamics potential and found that affected by logarithmic correction. Stability of the system using both specific heat and Hessian matrix analyzed and concluded that presence of the logarithmic correction may be helpful to have stability. By using the geometro-thermodynamic formalism we confirmed logarithmic corrected entropy for the charged dS black hole. Now, the logarithmic corrected entropy given by the equation (\ref{end}) can be used for the charged dS black hole to some various studies specially based on dS/CFT Correspondence. For example one can investigate $P$-$V$ criticality of the logarithmic corrected charged dS black hole. It may be also useful to study electrically charged de Sitter black hole in higher-derivative gravity \cite{higher}. Finally it would be interesting to consider logarithmic corrected charged dS black hole surrounded by quintessence to obtain extreme solutions \cite{quint}.

\end{document}